# Burst Eddy Current Testing with a Diamond Magnetometry

Chang Xu[1], Jixing Zhang[1*], Heng Yuan[1,2*], Guodong Bian[1], Pengcheng Fan[1], and Minxin Li[1].
[1]*School of Instrumentation and Optoelectronic Engineering, Beihang University, Beijing 100191, China;*
[2]*Research Institute of Frontier Science, Beihang University, Beijing 100191, China;*
*zhangjixing@buaa.edu.cn; hengyuan@buaa.edu.cn.

**Abstract**
In this work, a burst eddy current testing technique based on the employment of a diamond nitrogen vacancy (NV) center magnetometer with the Hahn echo (HE) sequence is demonstrated. With the confocal experiment apparatus, the HE-based NV magnetometer attained a magnetic sensitivity of $4.3\ \text{nT}/\sqrt{\text{Hz}}$ and a volume-normalized sensitivity of $3.6\ \text{pT}/\sqrt{\text{Hz}\cdot\text{mm}^{-3}}$, which are ~5 times better than the already existing method under the same conditions. Based on the proposed magnetometer configuration, a burst eddy current (BEC) testing prototype achieves a minimum detectable sample smaller than $300\ \mu\text{m}$ and measurement accuracy of $9.85\ \mu\text{m}$, which is employed to image different metallic specimens and detect the layered internal structures. Since our prototype comprises superb high sensitivity, it exhibits various potential applications in the fields of deformation monitoring, security screening, and quality control. Moreover, its biocompatibility and promising nanoscale resolution paves the way for electromagnetic testing in the fields of biomaterials.

Eddy current testing is widely applied for the imaging of conductive samples such as solutions[1,2] and metals[3], as well as for implementing non-destructive testing in quality control[4,5] and security screening[6]. More specifically, under an alternating current (AC) excitation magnetic field, the conductive samples generate near-surface eddy currents, while the associated secondary magnetic fields can be detected by magnetometers such as SQUID[7], GMR[8], and atomic magnetometers[9,10]. As far as the nitrogen vacancy (NV) centers in diamond are concerned, they have emerged as a versatile quantum spin system, enabling thus the execution of active research at the cutting edge of the quantum technologies, such as temperature sensing[11], electric field sensing[12] and biomedical imaging[13,14]. The enhanced sensitivity to the magnetic fields renders the NV centers suitable for magnetometers-based applications[15,16] with increased bandwidth[17] and sensitivity[18], which paves a new way for eddy current detection. The eddy current testing method by using an NV-based magnetometer has been already reported in the literature[19,20]. However, the existing NV eddy current testing approach is based on the implementation of a continuous-wave (CW) magnetometry scheme with insufficient sensitivity of $40\ \text{nT}/\sqrt{\text{Hz}}$ and bandwidth of $100\ \text{kHz}$, limiting thus the size and conductivity of the detectable objects. In addition, continuous eddy currents produce thermal effects, which hinders the wider application of the conventional scheme on temperature-sensitive samples.

Along these lines, in this work, a burst eddy current (BEC) testing technique with an NV magnetometer based on the Hahn echo (HE) sequence is proposed. An excitation waveform with a single cycle is called a burst, which produces the primary magnetic field applied to the samples. The HE sequence refocuses the dephasing induced by the inhomogenous static field, making the detection of AC signals more sensitive than the DC scheme[21]. More specifically, the proposed NV-BEC testing scheme attains a sensitivity of $4.3\ \text{nT}/\sqrt{\text{Hz}}$, which is ~10 times better than the reported NV imaging methods[19] over a wide frequency band from $100\ \text{kHz}$ to $3\ \text{MHz}$. Moreover, the corresponding volume-normalized sensitivity is $3.6\ \text{pT}/\sqrt{\text{Hz}\cdot\text{mm}^{-3}}$, which is ~5 times better than the existing method under the enforcement of the same conditions. With this NV magnetometer, a minimum detectable sample smaller than $300\ \mu\text{m}$ and measurement accuracy of $9.85\ \mu\text{m}$ were achieved for eddy current imaging of samples. In addition, the employed NV-BEC scheme applies only a burst magnetic field, which can avoid the manifestation of the adverse eddy current thermal effects on the material. By also considering the biocompatibility[22] and the nanoscale resolution perspective[16,23], this scheme is promising in order to extend the application of the eddy current testing to both biological materials, as well as microelectronic devices.

The experimental setup that is schematically illustrated in Fig. 1(a), consists of a diamond AC magnetometer operating in the HE mode, a BEC excitation coil, and the material under investigation. The diamond sample was fabricated by employing the chemical vapor deposition (CVD) method and using 50 ppm N impurity under $1\times10^{18}\ \text{e/cm}^2$ and 10 MeV electron irradiation. Subsequently, the sample was annealed for a total period of 2 h at 800 °C. The NV density is ~3 ppm. A permanent magnet was fixed on a translation stage with five degrees of freedom in order to generate the bias magnetic field for the diamond NV magnetometer. Additionally, a confocal scheme was used for the optical path[24]. The microwave was also applied to the NV centers through a ring-shaped antenna. The single-cycle excitation signal from an arbitrary waveform generator (Agilent 33522B) operating in the burst mode was enforced to a 3-mm-diameter coil with 18 turns fixed on the back of the diamond in order to generate the primary magnetic fields on both the diamond and the specimen. Furthermore, the specimens were placed on the end of a 20 cm insulated polyvinyl chloride rod whereas the other end was fixed to a motorized translation stage (Thorlabs PT3-Z8) in order to complete a point-by-point scan. The same experimental setup was also employed for examining the conventional continuous eddy current scheme for comparison.

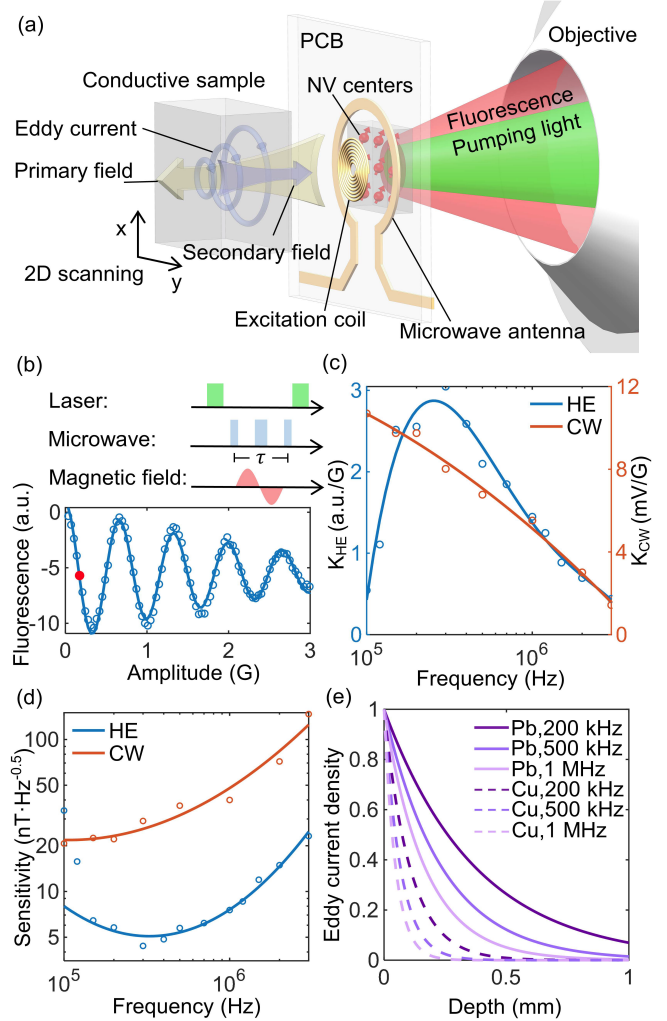

FIG. 1. (a) Schematic diagram of the eddy current excitation and detection probe, with the sample being tested. (b) The upper plot represents the sequence of the laser and the respective microwave pulses, along with the waveform of the excitation burst magnetic field. The lower plot indicates that the HE response of the excitation field varies periodically with the amplitude, using the curve of $y = ae^{-bx}\cos(cx) + d$ to fit the data points, in which $a$ represents the contrast, $b$ reflects the decoherence speed, $c$ is the spin oscillating angular velocity and $d$ is the bias. The red dot indicates one of the optimized working points. (c) Scale factor of the HE and CW methods for measuring the eddy currents. (d) Magnetic sensitivity of the HE and CW methods under the enforcement of different frequencies. (e) The density of generated eddy currents in various conductors decreases by increasing the depth into the sample at different excitation frequencies.

When the specimens were moved near the NV-BET testing probe, the AC magnetic field $B_p$ induces the eddy currents, producing thus a secondary magnetic field $B_S$ for reducing the total magnetic field amplitude and cause a phase shift according to Maxwell's equations. The fundamentals of the NV magnetic measurement process are based on the relationship between the ground state spin dynamics and the external magnetic field. More specifically, the NV center's ground state Hamiltonian contains the Zeeman splitting term $\gamma_e \vec{B} \cdot \vec{S}$, where $\gamma_e$ is the gyromagnetic ratio, $\vec{B}$ is the magnetic field, and $\vec{S}$ is the electronic spin operator. The induced energy level shift by this term can be measured by the optical detection magnetic resonance (ODMR) signal, providing the broadband DC magnetic measuring principle. NV magnetometer in the CW scheme with the lock-in amplifier is used in order to detect the change of the magnetic fields, which directly outputs the amplitude and phase shift. The corresponding eddy current scheme introduces a continuous AC magnetic field to the specimen, and the eddy currents cause the change of the total magnetic field, which is detected by the CW lock-in NV magnetometer. While in AC magnetic field sensing, the dynamical decoupling sequences can enhance the sensitivity of the NV centers to the AC magnetic field with a special frequency. Along these lines, in this work, the HE sequence is utilized for BEC testing, as it is divulged in Fig. 1(b). The laser pulses are employed for spin polarization and readout processes. The HE sequence consists of an array of microwave pulses, including a $\frac{\pi}{2}$ pulse, a $\pi$ pulse, and another $\frac{\pi}{2}$ pulse, as well as a sensing time $\tau$. Additionally, a single-cycle sinusoidal magnetic field with frequency $f = 1/\tau$ during the sensitivity time induces an accumulated phase $\phi = \int_0^{\tau/2} \gamma_e \left(B_S(t) + B_p(t)\right) dt + \int_{\tau/2}^{\tau} -\gamma_e \left(B_S(t) + B_p(t)\right) dt$, resulting in the change of the fluorescence signal of the NV centers. When the amplitude of the applied electrical signal is changed, the magnetic field changes accordingly and the fluorescence signal reveals a periodic alternating. The lower plot in Fig. 1(b) depicts the relationship between the amplitude of the excitation magnetic field and the response. The HE magnetometer exhibits a cosine response curve as a function of the linearly increased amplitude of the excitation magnetic field.

The scale factors $K$ at the optimal working point measured under different frequencies are presented in Fig. 1(c). The graph of the scale factors of the CW lock-in scheme under the same conditions is also shown. The HE scheme achieves the optimal parameters around 300 kHz. At lower applied frequencies, the scale factors become smaller due to the limited decoherence time ($T_2 \approx 6\ \mu s$), while at higher operating frequencies the scale factors decrease due to the shorter sensitive time and the less accumulated phase. The CW lock-in the scheme performs well at low frequencies, but the decrease of the frequency response of the NV center spin causes the scale factor to decrease at high frequencies.

The sensitivity of the HE measurement scheme $S$ is given by the following equation:

$$S = \frac{\Delta}{K}\sqrt{t}, \quad (1)$$

where $\Delta$ is the standard deviation and $t$ is the measurement time. The calculated sensitivities are illustrated in Fig. 1(d), where we can ascertain that the HE scheme exhibits the highest sensitivity at 300 kHz of 4.3 nT/$\sqrt{Hz}$. It is also worth pointing out that the proposed HE scheme possesses a higher sensitivity than the reported NV eddy current testing schemes[19, 20] in a wide frequency band. By considering that the detection volume is ~7×10$^{-7}$ mm$^3$, the volume-normalized sensitivity is 3.6 pT/$\sqrt{Hz \cdot mm^{-3}}$. On top of that, the samples with longer decoherence time have better scale factors at low frequencies, so the utilization of the diamond samples with high decoherence time[25] can lead to further improvement of the sensitivity. In addition, the enhancement of the polarization efficiency, the microwave power, the fluorescence collection efficiency[26], and the readout efficiency[27, 28] can also lead to further improvement of the sensitivity.

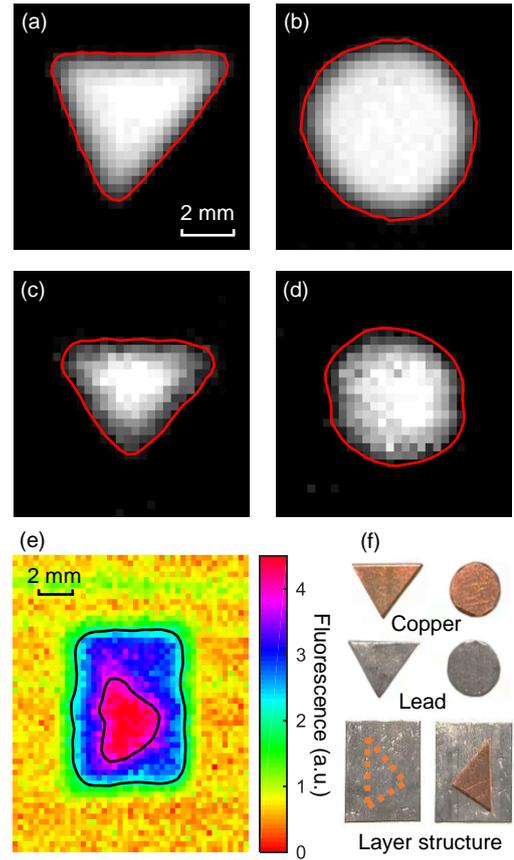

FIG.2. Depiction of the imaging results. (a)-(d) Direct imaging. The first row illustrates the eddy current imaging results of the copper, and the second row displays the imaging results of the lead. The red lines demonstrate the detected edge of the sample. (e) Penetration of the lead to image of the copper below. The color figure is the raw data taken at each point of the scan, whereas the black lines are the detected edge of the mask and target. (f) Photographs of the samples. The dashed line in the layer structure indicates the invisible copper under barriers, while next to it is the actual structure on the backside of the sample.

The imaging of the different metallic materials was accomplished with the previously described NV-BEC testing scheme. The circular and triangular copper (Cu) and lead (Pb) specimens (images in Fig. 2(f)) were scanned in two dimensions with a 300 kHz excitation signal and 0.3 mm step sizes. The fluorescence at each point has been recorded, and the corresponding imaging results are disclosed in Fig. 2(a-d). The extracted images clearly indicate that the existence of specimens, demonstrating a sub-mm imaging resolution. Furthermore, after performing the Gaussian smoothing and thresholding, the contours of the specimens are presented, which clearly indicate the edges of these shapes.

It is interesting to notice that the induced eddy current density decays with the increasing depth inside. More specifically, the depth at which the eddy current density decreases to the value of 1/e of the surface is called skin depth $\delta = 1/\sqrt{\pi f \mu \sigma}$, where $\mu$ is the permeability and $\sigma$ is the conductivity. Fig. 1(e) depicts the relationship between the eddy current density and the copper/lead specimens at different applied frequencies.

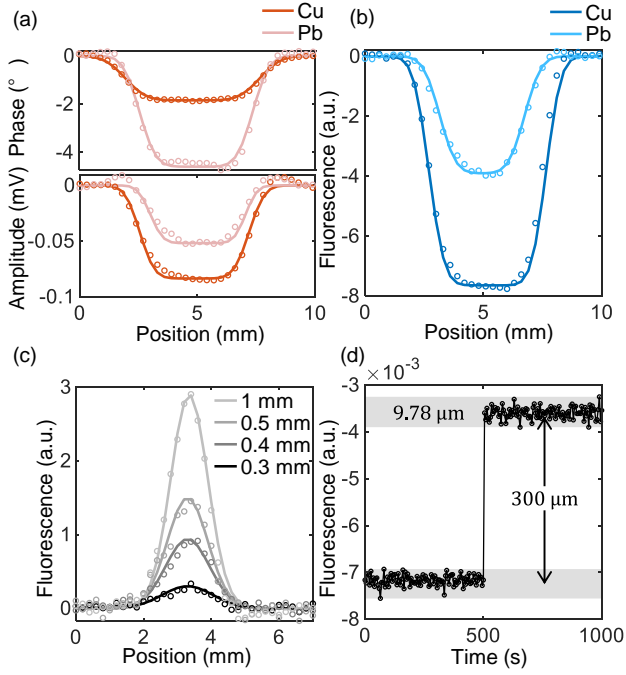

FIG.3. (a) Amplitude and phase response of the samples that were measured by the CW lock-in method by enforcing a one-dimensional scan. (b) Burst excitation response of the samples measured by HE sequence in a one-dimensional scan. (c) Scanning of ultra-small samples in order to measure the minimum detectable sample. (d) Depiction of a two-point scanning response obtained at the edge of the Cu sample with a maximum scale factor.

Since lead possesses lower conductivity than copper, its skin depth is significantly higher. Different materials' skin depths indicate that the eddy currents have the ability to penetrate conductive barriers and as a result detect hidden materials behind surfaces, permitting the nondestructive inspection of layer structures, as it is illustrated in Fig. 2(e). In the experiment, a triangular piece of copper was placed underneath the lead sheet to be masked, whereas the NV-BEC testing probe was used under the application of a frequency of 200 kHz instead of 300 kHz in order to penetrate the lead sheet with 0.3 mm step sizes. From the colored image, we can draw the conclusion that the lead skin and the hidden triangular copper structure underneath can be easily distinguished. After performing both Gaussian smoothing in combination with double threshold processing, the edges of the lead skin and the copper triangle can be clearly identified.

In order to qualitatively compare the performance of the burst and the conventional CW schemes, one-dimensional scanning experiments were carried out in both a conventional CW lock-in scheme and a BEC scheme. Copper (10 × 7 × 0.5 mm) and lead (10 × 7 × 0.4 mm) sheets were used for demonstrating the universality of the imaging schemes. In addition, one-dimensional scans were performed for each of the two metal sheets under the enforcement of a 300 kHz excitation signal with 0.3 mm step sizes, and all data were normalized to zero at the insulated position. The results are divulged in Fig. 3(a-b), in which the data points are fitted as a convolution of a Gaussian function with a square wave. As far as the CW lock-in scheme is concerned, the lead amplitude response is smaller and the phase response is bigger, while for the copper a reduced phase response and an elevated amplitude response were recorded. The BEC scheme provides stable measurements for both metals, while the response is larger for the object with the higher conductivity. The scheme exhibits a steep slope, which is superior to other measurement schemes, such as the cold atomic magnetometer with higher magnetic sensitivity[29].

In order to further investigate the minimum detectable sample size of this NV-BEC testing scheme, a one-dimensional imaging was performed on small samples with frequency value of $f$ = 300 kHz. Fig. 3(c) illustrates the scan results for samples of different sizes. The data points are fitted by using a Gaussian function. The plot reveals that the response decreases with a smaller width, and still presents a clear signal for the sample of 0.3 mm diameter, indicating that this NV-BEC probe is able to detect a minimum sample size smaller than 0.3 mm, which is better than the smallest samples of the reported scheme[20, 29]. It should be noted that the implementation of a coil's 3 mm diameter, instead of the NV magnetometer spatial resolution, limits the minimum detectable size. A multi-turn sub-mm excitation coil, which can produce larger excitation field with a smaller size, results in a smaller detectable sample, as well as a better spatial resolution.

Finally, the measurement accuracy of the scheme is thoroughly analyzed. By operating at a frequency value of 300 kHz, and enforcing scanning in a range $\Delta x = 300$ μm, each point was averaged in 5 s in order to obtain the results of Fig. 3(d), yielding an average contrast change $\Delta c = 3.6 \times 10^{-3}$. The scale factor at this point can be calculated by using the following expression: $K = \frac{\Delta c}{\Delta x} = 1.19 \times 10^{-5}$ /μm. Besides, the standard deviation of the two sets of data is $\Delta = 1.17 \times 10^{-4}$. Thus, the measurement accuracy $u$ can be calculated as $u = \frac{\Delta}{K} = 9.85$ μm.

In this work, we present a burst eddy current testing technique based on a diamond AC magnetometer with the HE sequence of NV centers. This scheme is used for the imaging of different conductive materials, as well as for the detection through conductive masks. In addition, from our approach, a sensitivity of $4.3 \text{ nT}/\sqrt{\text{Hz}}$, as well as a volume-normalized sensitivity of $3.6 \text{ pT}/\sqrt{\text{Hz} \cdot \text{mm}^{-3}}$ from 100 kHz to 3 MHz is achieved. The imaging system attains also a minimum detectable sample smaller than 300 μm and measurement accuracy of 9.85 μm. Based on the proposed approach, versatile applications could emerge including the identification and the classification of various targets combined with the machine learning techniques[30], real-time deformation monitoring, security and quality control with the further promotion on integrated[31] or on-chip[32] NV magnetometry technologies. The proposed technology also takes one step further for the prospect of the eddy current testing scheme to various fields of biological tissues and microelectronics[33].